\documentclass[longtitle,preprint]{elsarticle}

\usepackage{graphicx}
\usepackage{rotating}

\begin{document}

\newcommand{\fix}[1]{} 
\newcommand{\MET}{\mbox{$\raisebox{.3ex}{$\not$}E_{\rm T}$}}
\def\DZero{D\O~}
\def\GeV  {\ensuremath{\mathrm{ Ge\kern -0.1em V } }}
\def\GeVc2{\ensuremath{\mathrm{ Ge\kern -0.1em V }\kern -0.2em /c^2 }}
\setlength{\topmargin}{.0cm}

\renewcommand\theaffn{\arabic{affn}}
\renewcommand\thefootnote{\fnsymbol{footnote}}



\begin{frontmatter}

\title{Measurement of $W$-Boson Helicity Fractions \\
   in Top-Quark Decays Using $\cos\theta^*$}
 
\vspace{0.5in}

\date{\today}

\author[DivisionHighEnergyPhysicsDepartmentPhysicsHelsinkiandHelsinkiPhysicsFIN00014HelsinkiFinland]{T.~Aaltonen}
\author[EnricoFermiChicagoChicagoIllinois60637]{J.~Adelman}
\author[TsukubaTsukubaIbaraki305Japan]{T.~Akimoto}
\author[InstitutodeFisicadeCantabriaCSICCantabria39005SantanderSpain]{B.~\'{A}lvarez~Gonz\'{a}lez$^t$}
\author[IstitutoNazionalediFisicaNucleareSezionediPadovaTrentozPadovaI35131PadovaItaly]{S.~Amerio$^z$}
\author[MichiganAnnArborMichigan48109]{D.~Amidei}
\author[NorthwesternEvanstonIllinois60208]{A.~Anastassov}
\author[LaboratoriNazionalidiFrascatiIstitutoNazionalediFisicaNucleareI00044FrascatiItaly]{A.~Annovi}
\author[Comenius84248BratislavaSlovakiaExperimentalPhysics04001KosiceSlovakia]{J.~Antos}
\author[FermiNationalAcceleratorLaboratoryBataviaIllinois60510]{G.~Apollinari}
\author[PurdueWestLafayetteIndiana47907]{A.~Apresyan}
\author[WasedaTokyo169Japan]{T.~Arisawa}
\author[JointforNuclearResearchRU141980DubnaRussia]{A.~Artikov}
\author[FermiNationalAcceleratorLaboratoryBataviaIllinois60510]{W.~Ashmanskas}
\author[InstitutdeFisicadAltesEnergiesUniversitatAutonomadeBarcelonaE08193Bellaterra(Barcelona)Spain]{A.~Attal}
\author[TexasAMCollegeStationTexas77843]{A.~Aurisano}
\author[OxfordOxfordOX13RHUnitedKingdom]{F.~Azfar}
\author[FermiNationalAcceleratorLaboratoryBataviaIllinois60510]{W.~Badgett}
\author[ErnestOrlandoLawrenceBerkeleyNationalLaboratoryBerkeleyCalifornia94720]{A.~Barbaro-Galtieri}
\author[PurdueWestLafayetteIndiana47907]{V.E.~Barnes}
\author[TheJohnsHopkinsBaltimoreMaryland21218]{B.A.~Barnett}
\author[IstitutoNazionalediFisicaNuclearePisaaaPisabbSienaandccScuolaNormaleSuperioreI56127PisaItaly]{P.~Barria$^{bb}$}
\author[CollegeLondonLondonWC1E6BTUnitedKingdom]{V.~Bartsch}
\author[MassachusettsTechnologyCambridgeMassachusetts02139]{G.~Bauer}
\author[ParticlePhysicsMcGillMontrealQuebecCanadaH3A2T8SimonFraserBurnabyBritishColumbiaCanadaV5A1S6TorontoTorontoOntarioCanadaM5S1A7andTRIUMFVancouverBritishColumbiaCanadaV6T2A3]{P.-H.~Beauchemin}
\author[IstitutoNazionalediFisicaNuclearePisaaaPisabbSienaandccScuolaNormaleSuperioreI56127PisaItaly]{F.~Bedeschi}
\author[CollegeLondonLondonWC1E6BTUnitedKingdom]{D.~Beecher}
\author[TheJohnsHopkinsBaltimoreMaryland21218]{S.~Behari}
\author[IstitutoNazionalediFisicaNuclearePisaaaPisabbSienaandccScuolaNormaleSuperioreI56127PisaItaly]{G.~Bellettini$^{aa}$}
\author[WisconsinMadisonWisconsin53706]{J.~Bellinger}
\author[DukeDurhamNorthCarolina27708]{D.~Benjamin}
\author[FermiNationalAcceleratorLaboratoryBataviaIllinois60510]{A.~Beretvas}
\author[ErnestOrlandoLawrenceBerkeleyNationalLaboratoryBerkeleyCalifornia94720]{J.~Beringer}
\author[TheRockefellerNewYorkNewYork10021]{A.~Bhatti}
\author[FermiNationalAcceleratorLaboratoryBataviaIllinois60510]{M.~Binkley}
\author[IstitutoNazionalediFisicaNucleareSezionediPadovaTrentozPadovaI35131PadovaItaly]{D.~Bisello$^z$}
\author[CollegeLondonLondonWC1E6BTUnitedKingdom]{I.~Bizjak$^{ff}$}
\author[ArgonneNationalLaboratoryArgonneIllinois60439]{R.E.~Blair}
\author[BrandeisWalthamMassachusetts02254]{C.~Blocker}
\author[TheJohnsHopkinsBaltimoreMaryland21218]{B.~Blumenfeld}
\author[DukeDurhamNorthCarolina27708]{A.~Bocci}
\author[RochesterRochesterNewYork14627]{A.~Bodek}
\author[RochesterRochesterNewYork14627]{V.~Boisvert}
\author[PurdueWestLafayetteIndiana47907]{G.~Bolla}
\author[PurdueWestLafayetteIndiana47907]{D.~Bortoletto}
\author[PittsburghPittsburghPennsylvania15260]{J.~Boudreau}
\author[CaliforniaSantaBarbaraSantaBarbaraCalifornia93106]{A.~Boveia}
\author[CaliforniaSantaBarbaraSantaBarbaraCalifornia93106]{B.~Brau$^a$}
\author[IllinoisUrbanaIllinois61801]{A.~Bridgeman}
\author[IstitutoNazionalediFisicaNucleareBolognayBolognaI40127BolognaItaly]{L.~Brigliadori$^y$}
\author[MichiganStateEastLansingMichigan48824]{C.~Bromberg}
\author[EnricoFermiChicagoChicagoIllinois60637]{E.~Brubaker}
\author[JointforNuclearResearchRU141980DubnaRussia]{J.~Budagov}
\author[RochesterRochesterNewYork14627]{H.S.~Budd}
\author[IllinoisUrbanaIllinois61801]{S.~Budd}
\author[FermiNationalAcceleratorLaboratoryBataviaIllinois60510]{S.~Burke}
\author[FermiNationalAcceleratorLaboratoryBataviaIllinois60510]{K.~Burkett}
\author[IstitutoNazionalediFisicaNucleareSezionediPadovaTrentozPadovaI35131PadovaItaly]{G.~Busetto$^z$}
\author[GlasgowGlasgowG128QQUnitedKingdom]{P.~Bussey}
\author[ParticlePhysicsMcGillMontrealQuebecCanadaH3A2T8SimonFraserBurnabyBritishColumbiaCanadaV5A1S6TorontoTorontoOntarioCanadaM5S1A7andTRIUMFVancouverBritishColumbiaCanadaV6T2A3]{A.~Buzatu}
\author[ArgonneNationalLaboratoryArgonneIllinois60439]{K.~L.~Byrum}
\author[DukeDurhamNorthCarolina27708]{S.~Cabrera$^v$}
\author[CentrodeInvestigacionesEnergeticasMedioambientalesyTecnologicasE28040MadridSpain]{C.~Calancha}
\author[MichiganStateEastLansingMichigan48824]{M.~Campanelli}
\author[MichiganAnnArborMichigan48109]{M.~Campbell}
\author[FermiNationalAcceleratorLaboratoryBataviaIllinois60510]{F.~Canelli$^{14}$}
\author[PennsylvaniaPhiladelphiaPennsylvania19104]{A.~Canepa}
\author[IllinoisUrbanaIllinois61801]{B.~Carls}
\author[WisconsinMadisonWisconsin53706]{D.~Carlsmith}
\author[IstitutoNazionalediFisicaNuclearePisaaaPisabbSienaandccScuolaNormaleSuperioreI56127PisaItaly]{R.~Carosi}
\author[FloridaGainesvilleFlorida32611]{S.~Carrillo$^n$}
\author[ParticlePhysicsMcGillMontrealQuebecCanadaH3A2T8SimonFraserBurnabyBritishColumbiaCanadaV5A1S6TorontoTorontoOntarioCanadaM5S1A7andTRIUMFVancouverBritishColumbiaCanadaV6T2A3]{S.~Carron}
\author[InstitutodeFisicadeCantabriaCSICCantabria39005SantanderSpain]{B.~Casal}
\author[FermiNationalAcceleratorLaboratoryBataviaIllinois60510]{M.~Casarsa}
\author[IstitutoNazionalediFisicaNucleareBolognayBolognaI40127BolognaItaly]{A.~Castro$^y$}
\author[IstitutoNazionalediFisicaNuclearePisaaaPisabbSienaandccScuolaNormaleSuperioreI56127PisaItaly]{P.~Catastini$^{bb}$}
\author[IstitutoNazionalediFisicaNucleareTriesteUdineI34100TriesteeeTriesteUdineI33100UdineItaly]{D.~Cauz$^{ee}$}
\author[IstitutoNazionalediFisicaNuclearePisaaaPisabbSienaandccScuolaNormaleSuperioreI56127PisaItaly]{V.~Cavaliere$^{bb}$}
\author[InstitutdeFisicadAltesEnergiesUniversitatAutonomadeBarcelonaE08193Bellaterra(Barcelona)Spain]{M.~Cavalli-Sforza}
\author[ErnestOrlandoLawrenceBerkeleyNationalLaboratoryBerkeleyCalifornia94720]{A.~Cerri}
\author[CollegeLondonLondonWC1E6BTUnitedKingdom]{L.~Cerrito$^p$}
\author[CenterforHighEnergyPhysicsKyungpookNationalDaegu702701KoreaSeoulNationalSeoul151742KoreaSungkyunkwanSuwon440746KoreaKoreaScienceandTechnologyInformationDaejeon305806KoreaChonnamNationalGwangju500757Korea]{S.H.~Chang}
\author[PhysicsAcademiaSinicaTaipeiTaiwan11529RepublicChina]{Y.C.~Chen}
\author[CaliforniaDavisDavisCalifornia95616]{M.~Chertok}
\author[IstitutoNazionalediFisicaNuclearePisaaaPisabbSienaandccScuolaNormaleSuperioreI56127PisaItaly]{G.~Chiarelli}
\author[FermiNationalAcceleratorLaboratoryBataviaIllinois60510]{G.~Chlachidze}
\author[FermiNationalAcceleratorLaboratoryBataviaIllinois60510]{F.~Chlebana}
\author[CenterforHighEnergyPhysicsKyungpookNationalDaegu702701KoreaSeoulNationalSeoul151742KoreaSungkyunkwanSuwon440746KoreaKoreaScienceandTechnologyInformationDaejeon305806KoreaChonnamNationalGwangju500757Korea]{K.~Cho}
\author[JointforNuclearResearchRU141980DubnaRussia]{D.~Chokheli}
\author[HarvardCambridgeMassachusetts02138]{J.P.~Chou}
\author[MassachusettsTechnologyCambridgeMassachusetts02139]{G.~Choudalakis}
\author[RutgersPiscatawayNewJersey08855]{S.H.~Chuang}
\author[CarnegieMellonPittsburghPA15213]{K.~Chung}
\author[WisconsinMadisonWisconsin53706]{W.H.~Chung}
\author[RochesterRochesterNewYork14627]{Y.S.~Chung}
\author[Institutf"urExperimentelleKernphysikUniversit"atKarlsruhe76128KarlsruheGermany]{T.~Chwalek}
\author[LPNHEUniversitePierreetMarieCurieIN2P3CNRSUMR7585ParisF75252France]{C.I.~Ciobanu}
\author[IstitutoNazionalediFisicaNuclearePisaaaPisabbSienaandccScuolaNormaleSuperioreI56127PisaItaly]{M.A.~Ciocci$^{bb}$}
% ...
\author[CenterforHighEnergyPhysicsKyungpookNationalDaegu702701KoreaSeoulNationalSeoul151742KoreaSungkyunkwanSuwon440746KoreaKoreaScienceandTechnologyInformationDaejeon305806KoreaChonnamNationalGwangju500757Korea]{E.J.~Jeon}
\author[IstitutoNazionalediFisicaNucleareBolognayBolognaI40127BolognaItaly]{M.K.~Jha}
\author[FermiNationalAcceleratorLaboratoryBataviaIllinois60510]{S.~Jindariani}
\author[CaliforniaDavisDavisCalifornia95616]{W.~Johnson}
\author[PurdueWestLafayetteIndiana47907]{M.~Jones}
\author[CenterforHighEnergyPhysicsKyungpookNationalDaegu702701KoreaSeoulNationalSeoul151742KoreaSungkyunkwanSuwon440746KoreaKoreaScienceandTechnologyInformationDaejeon305806KoreaChonnamNationalGwangju500757Korea]{K.K.~Joo}
\author[CarnegieMellonPittsburghPA15213]{S.Y.~Jun}
\author[CenterforHighEnergyPhysicsKyungpookNationalDaegu702701KoreaSeoulNationalSeoul151742KoreaSungkyunkwanSuwon440746KoreaKoreaScienceandTechnologyInformationDaejeon305806KoreaChonnamNationalGwangju500757Korea]{J.E.~Jung}
\author[FermiNationalAcceleratorLaboratoryBataviaIllinois60510]{T.R.~Junk}
\author[TexasAMCollegeStationTexas77843]{T.~Kamon}
\author[FloridaGainesvilleFlorida32611]{D.~Kar}
\author[WayneStateDetroitMichigan48201]{P.E.~Karchin}
\author[OsakaCityOsaka588Japan]{Y.~Kato$^l$}
\author[FermiNationalAcceleratorLaboratoryBataviaIllinois60510]{R.~Kephart}
\author[EnricoFermiChicagoChicagoIllinois60637]{W.~Ketchum}
\author[PennsylvaniaPhiladelphiaPennsylvania19104]{J.~Keung}
\author[TexasAMCollegeStationTexas77843]{V.~Khotilovich}
\author[FermiNationalAcceleratorLaboratoryBataviaIllinois60510]{B.~Kilminster}
\author[CenterforHighEnergyPhysicsKyungpookNationalDaegu702701KoreaSeoulNationalSeoul151742KoreaSungkyunkwanSuwon440746KoreaKoreaScienceandTechnologyInformationDaejeon305806KoreaChonnamNationalGwangju500757Korea]{D.H.~Kim}
\author[CenterforHighEnergyPhysicsKyungpookNationalDaegu702701KoreaSeoulNationalSeoul151742KoreaSungkyunkwanSuwon440746KoreaKoreaScienceandTechnologyInformationDaejeon305806KoreaChonnamNationalGwangju500757Korea]{H.S.~Kim}
\author[CenterforHighEnergyPhysicsKyungpookNationalDaegu702701KoreaSeoulNationalSeoul151742KoreaSungkyunkwanSuwon440746KoreaKoreaScienceandTechnologyInformationDaejeon305806KoreaChonnamNationalGwangju500757Korea]{H.W.~Kim}
\author[CenterforHighEnergyPhysicsKyungpookNationalDaegu702701KoreaSeoulNationalSeoul151742KoreaSungkyunkwanSuwon440746KoreaKoreaScienceandTechnologyInformationDaejeon305806KoreaChonnamNationalGwangju500757Korea]{J.E.~Kim}
\author[LaboratoriNazionalidiFrascatiIstitutoNazionalediFisicaNucleareI00044FrascatiItaly]{M.J.~Kim}
\author[CenterforHighEnergyPhysicsKyungpookNationalDaegu702701KoreaSeoulNationalSeoul151742KoreaSungkyunkwanSuwon440746KoreaKoreaScienceandTechnologyInformationDaejeon305806KoreaChonnamNationalGwangju500757Korea]{S.B.~Kim}
\author[TsukubaTsukubaIbaraki305Japan]{S.H.~Kim}
\author[EnricoFermiChicagoChicagoIllinois60637]{Y.K.~Kim}
\author[TsukubaTsukubaIbaraki305Japan]{N.~Kimura}
\author[BrandeisWalthamMassachusetts02254]{L.~Kirsch}
\author[FloridaGainesvilleFlorida32611]{S.~Klimenko}
\author[MassachusettsTechnologyCambridgeMassachusetts02139]{B.~Knuteson}
\author[DukeDurhamNorthCarolina27708]{B.R.~Ko}
\author[WasedaTokyo169Japan]{K.~Kondo}
\author[CenterforHighEnergyPhysicsKyungpookNationalDaegu702701KoreaSeoulNationalSeoul151742KoreaSungkyunkwanSuwon440746KoreaKoreaScienceandTechnologyInformationDaejeon305806KoreaChonnamNationalGwangju500757Korea]{D.J.~Kong}
\author[FloridaGainesvilleFlorida32611]{J.~Konigsberg}
\author[FloridaGainesvilleFlorida32611]{A.~Korytov}
\author[DukeDurhamNorthCarolina27708]{A.V.~Kotwal}
\author[Institutf"urExperimentelleKernphysikUniversit"atKarlsruhe76128KarlsruheGermany]{M.~Kreps}
\author[PennsylvaniaPhiladelphiaPennsylvania19104]{J.~Kroll}
\author[EnricoFermiChicagoChicagoIllinois60637]{D.~Krop}
\author[BaylorWacoTexas76798]{N.~Krumnack}
\author[DukeDurhamNorthCarolina27708]{M.~Kruse}
\author[CaliforniaSantaBarbaraSantaBarbaraCalifornia93106]{V.~Krutelyov}
\author[TsukubaTsukubaIbaraki305Japan]{T.~Kubo}
\author[Institutf"urExperimentelleKernphysikUniversit"atKarlsruhe76128KarlsruheGermany]{T.~Kuhr}
\author[WayneStateDetroitMichigan48201]{N.P.~Kulkarni}
\author[TsukubaTsukubaIbaraki305Japan]{M.~Kurata}
\author[EnricoFermiChicagoChicagoIllinois60637]{S.~Kwang}
\author[PurdueWestLafayetteIndiana47907]{A.T.~Laasanen}
\author[IstitutoNazionalediFisicaNuclearePisaaaPisabbSienaandccScuolaNormaleSuperioreI56127PisaItaly]{S.~Lami}
\author[FermiNationalAcceleratorLaboratoryBataviaIllinois60510]{S.~Lammel}
\author[CollegeLondonLondonWC1E6BTUnitedKingdom]{M.~Lancaster}
\author[CaliforniaDavisDavisCalifornia95616]{R.L.~Lander}
\author[TheOhioStateColumbusOhio43210]{K.~Lannon$^s$}
\author[RutgersPiscatawayNewJersey08855]{A.~Lath}
\author[IstitutoNazionalediFisicaNuclearePisaaaPisabbSienaandccScuolaNormaleSuperioreI56127PisaItaly]{G.~Latino$^{bb}$}
\author[IstitutoNazionalediFisicaNucleareSezionediPadovaTrentozPadovaI35131PadovaItaly]{I.~Lazzizzera$^z$}
\author[ArgonneNationalLaboratoryArgonneIllinois60439]{T.~LeCompte}
\author[TexasAMCollegeStationTexas77843]{E.~Lee}
\author[EnricoFermiChicagoChicagoIllinois60637]{H.S.~Lee}
\author[TexasAMCollegeStationTexas77843]{S.W.~Lee$^u$}
\author[IstitutoNazionalediFisicaNuclearePisaaaPisabbSienaandccScuolaNormaleSuperioreI56127PisaItaly]{S.~Leone}
\author[FermiNationalAcceleratorLaboratoryBataviaIllinois60510]{J.D.~Lewis}
\author[ErnestOrlandoLawrenceBerkeleyNationalLaboratoryBerkeleyCalifornia94720]{C.-S.~Lin}
\author[OxfordOxfordOX13RHUnitedKingdom]{J.~Linacre}
\author[FermiNationalAcceleratorLaboratoryBataviaIllinois60510]{M.~Lindgren}
\author[PennsylvaniaPhiladelphiaPennsylvania19104]{E.~Lipeles}
\author[CaliforniaDavisDavisCalifornia95616]{A.~Lister}
\author[FermiNationalAcceleratorLaboratoryBataviaIllinois60510]{D.O.~Litvintsev}
\author[PittsburghPittsburghPennsylvania15260]{C.~Liu}
\author[FermiNationalAcceleratorLaboratoryBataviaIllinois60510]{T.~Liu}
\author[PennsylvaniaPhiladelphiaPennsylvania19104]{N.S.~Lockyer}
\author[YaleNewHavenConnecticut06520]{A.~Loginov}
\author[IstitutoNazionalediFisicaNucleareSezionediPadovaTrentozPadovaI35131PadovaItaly]{M.~Loreti$^z$}
\author[Comenius84248BratislavaSlovakiaExperimentalPhysics04001KosiceSlovakia]{L.~Lovas}
\author[IstitutoNazionalediFisicaNucleareSezionediPadovaTrentozPadovaI35131PadovaItaly]{D.~Lucchesi$^z$}
\author[IstitutoNazionalediFisicaNucleareSezionediRoma1ddSapienzaUniversitadiRomaI00185RomaItaly]{C.~Luci$^{dd}$}
\author[Institutf"urExperimentelleKernphysikUniversit"atKarlsruhe76128KarlsruheGermany]{J.~Lueck}
\author[ErnestOrlandoLawrenceBerkeleyNationalLaboratoryBerkeleyCalifornia94720]{P.~Lujan}
\author[FermiNationalAcceleratorLaboratoryBataviaIllinois60510]{P.~Lukens}
\author[TheRockefellerNewYorkNewYork10021]{G.~Lungu}
\author[OxfordOxfordOX13RHUnitedKingdom]{L.~Lyons}
\author[ErnestOrlandoLawrenceBerkeleyNationalLaboratoryBerkeleyCalifornia94720]{J.~Lys}
\author[Comenius84248BratislavaSlovakiaExperimentalPhysics04001KosiceSlovakia]{R.~Lysak}
\author[ParticlePhysicsMcGillMontrealQuebecCanadaH3A2T8SimonFraserBurnabyBritishColumbiaCanadaV5A1S6TorontoTorontoOntarioCanadaM5S1A7andTRIUMFVancouverBritishColumbiaCanadaV6T2A3]{D.~MacQueen}
\author[FermiNationalAcceleratorLaboratoryBataviaIllinois60510]{R.~Madrak}
\author[FermiNationalAcceleratorLaboratoryBataviaIllinois60510]{K.~Maeshima}
\author[MassachusettsTechnologyCambridgeMassachusetts02139]{K.~Makhoul}
\author[DivisionHighEnergyPhysicsDepartmentPhysicsHelsinkiandHelsinkiPhysicsFIN00014HelsinkiFinland]{T.~Maki}
\author[TheJohnsHopkinsBaltimoreMaryland21218]{P.~Maksimovic}
\author[OxfordOxfordOX13RHUnitedKingdom]{S.~Malde}
\author[CollegeLondonLondonWC1E6BTUnitedKingdom]{S.~Malik}
\author[LiverpoolLiverpoolL697ZEUnitedKingdom]{G.~Manca$^e$}
\author[Athens15771AthensGreece]{A.~Manousakis-Katsikakis}
\author[PurdueWestLafayetteIndiana47907]{F.~Margaroli}
\author[Institutf"urExperimentelleKernphysikUniversit"atKarlsruhe76128KarlsruheGermany]{C.~Marino}
\author[IllinoisUrbanaIllinois61801]{C.P.~Marino}
\author[YaleNewHavenConnecticut06520]{A.~Martin}
\author[GlasgowGlasgowG128QQUnitedKingdom]{V.~Martin$^k$}
\author[InstitutdeFisicadAltesEnergiesUniversitatAutonomadeBarcelonaE08193Bellaterra(Barcelona)Spain]{M.~Mart\'{\i}nez}
\author[CentrodeInvestigacionesEnergeticasMedioambientalesyTecnologicasE28040MadridSpain]{R.~Mart\'{\i}nez-Ballar\'{\i}n}
\author[TsukubaTsukubaIbaraki305Japan]{T.~Maruyama}
\author[IstitutoNazionalediFisicaNucleareSezionediRoma1ddSapienzaUniversitadiRomaI00185RomaItaly]{P.~Mastrandrea}
\author[TsukubaTsukubaIbaraki305Japan]{T.~Masubuchi}
\author[TheJohnsHopkinsBaltimoreMaryland21218]{M.~Mathis}
\author[WayneStateDetroitMichigan48201]{M.E.~Mattson}
\author[IstitutoNazionalediFisicaNucleareBolognayBolognaI40127BolognaItaly]{P.~Mazzanti}
\author[RochesterRochesterNewYork14627]{K.S.~McFarland}
\author[TexasAMCollegeStationTexas77843]{P.~McIntyre}
\author[LiverpoolLiverpoolL697ZEUnitedKingdom]{R.~McNulty$^j$}
\author[LiverpoolLiverpoolL697ZEUnitedKingdom]{A.~Mehta}
\author[DivisionHighEnergyPhysicsDepartmentPhysicsHelsinkiandHelsinkiPhysicsFIN00014HelsinkiFinland]{P.~Mehtala}
\author[IstitutoNazionalediFisicaNuclearePisaaaPisabbSienaandccScuolaNormaleSuperioreI56127PisaItaly]{A.~Menzione}
\author[PurdueWestLafayetteIndiana47907]{P.~Merkel}
\author[TheRockefellerNewYorkNewYork10021]{C.~Mesropian}
\author[FermiNationalAcceleratorLaboratoryBataviaIllinois60510]{T.~Miao}
\author[BrandeisWalthamMassachusetts02254]{N.~Miladinovic}
\author[MichiganStateEastLansingMichigan48824]{R.~Miller}
\author[HarvardCambridgeMassachusetts02138]{C.~Mills}
\author[Institutf"urExperimentelleKernphysikUniversit"atKarlsruhe76128KarlsruheGermany]{M.~Milnik}
\author[PhysicsAcademiaSinicaTaipeiTaiwan11529RepublicChina]{A.~Mitra}
\author[FloridaGainesvilleFlorida32611]{G.~Mitselmakher}
\author[TsukubaTsukubaIbaraki305Japan]{H.~Miyake}
\author[HarvardCambridgeMassachusetts02138]{S.~Moed}
\author[IstitutoNazionalediFisicaNucleareBolognayBolognaI40127BolognaItaly]{N.~Moggi}
\author[CenterforHighEnergyPhysicsKyungpookNationalDaegu702701KoreaSeoulNationalSeoul151742KoreaSungkyunkwanSuwon440746KoreaKoreaScienceandTechnologyInformationDaejeon305806KoreaChonnamNationalGwangju500757Korea]{C.S.~Moon}
\author[FermiNationalAcceleratorLaboratoryBataviaIllinois60510]{R.~Moore}
\author[IstitutoNazionalediFisicaNuclearePisaaaPisabbSienaandccScuolaNormaleSuperioreI56127PisaItaly]{M.J.~Morello}
\author[Institutf"urExperimentelleKernphysikUniversit"atKarlsruhe76128KarlsruheGermany]{J.~Morlock}
\author[FermiNationalAcceleratorLaboratoryBataviaIllinois60510]{P.~Movilla~Fernandez}
\author[ErnestOrlandoLawrenceBerkeleyNationalLaboratoryBerkeleyCalifornia94720]{J.~M\"ulmenst\"adt}
\author[FermiNationalAcceleratorLaboratoryBataviaIllinois60510]{A.~Mukherjee}
\author[Institutf"urExperimentelleKernphysikUniversit"atKarlsruhe76128KarlsruheGermany]{Th.~Muller}
\author[TheJohnsHopkinsBaltimoreMaryland21218]{R.~Mumford}
\author[FermiNationalAcceleratorLaboratoryBataviaIllinois60510]{P.~Murat}
\author[IstitutoNazionalediFisicaNucleareBolognayBolognaI40127BolognaItaly]{M.~Mussini$^y$}
\author[FermiNationalAcceleratorLaboratoryBataviaIllinois60510]{J.~Nachtman$^o$}
\author[TsukubaTsukubaIbaraki305Japan]{Y.~Nagai}
\author[TsukubaTsukubaIbaraki305Japan]{A.~Nagano}
\author[TsukubaTsukubaIbaraki305Japan]{J.~Naganoma}
\author[TsukubaTsukubaIbaraki305Japan]{K.~Nakamura}
\author[OkayamaOkayama7008530Japan]{I.~Nakano}
\author[TuftsMedfordMassachusetts02155]{A.~Napier}
\author[DukeDurhamNorthCarolina27708]{V.~Necula}
\author[WisconsinMadisonWisconsin53706]{J.~Nett}
\author[PennsylvaniaPhiladelphiaPennsylvania19104]{C.~Neu$^w$}
\author[IllinoisUrbanaIllinois61801]{M.S.~Neubauer}
\author[Institutf"urExperimentelleKernphysikUniversit"atKarlsruhe76128KarlsruheGermany]{S.~Neubauer}
\author[ErnestOrlandoLawrenceBerkeleyNationalLaboratoryBerkeleyCalifornia94720]{J.~Nielsen$^g$}
\author[ArgonneNationalLaboratoryArgonneIllinois60439]{L.~Nodulman}
\author[CaliforniaSanDiegoLaJollaCalifornia92093]{M.~Norman}
\author[IllinoisUrbanaIllinois61801]{O.~Norniella}
\author[CollegeLondonLondonWC1E6BTUnitedKingdom]{E.~Nurse}
\author[OxfordOxfordOX13RHUnitedKingdom]{L.~Oakes}
\author[DukeDurhamNorthCarolina27708]{S.H.~Oh}
\author[CenterforHighEnergyPhysicsKyungpookNationalDaegu702701KoreaSeoulNationalSeoul151742KoreaSungkyunkwanSuwon440746KoreaKoreaScienceandTechnologyInformationDaejeon305806KoreaChonnamNationalGwangju500757Korea]{Y.D.~Oh}
\author[FloridaGainesvilleFlorida32611]{I.~Oksuzian}
\author[OsakaCityOsaka588Japan]{T.~Okusawa}
\author[DivisionHighEnergyPhysicsDepartmentPhysicsHelsinkiandHelsinkiPhysicsFIN00014HelsinkiFinland]{R.~Orava}
\author[DivisionHighEnergyPhysicsDepartmentPhysicsHelsinkiandHelsinkiPhysicsFIN00014HelsinkiFinland]{K.~Osterberg}
\author[IstitutoNazionalediFisicaNucleareSezionediPadovaTrentozPadovaI35131PadovaItaly]{S.~Pagan~Griso$^z$}
\author[FermiNationalAcceleratorLaboratoryBataviaIllinois60510]{E.~Palencia}
\author[FermiNationalAcceleratorLaboratoryBataviaIllinois60510]{V.~Papadimitriou}
\author[Institutf"urExperimentelleKernphysikUniversit"atKarlsruhe76128KarlsruheGermany]{A.~Papaikonomou}
\author[EnricoFermiChicagoChicagoIllinois60637]{A.A.~Paramonov}
\author[TheOhioStateColumbusOhio43210]{B.~Parks}
% ...
\author[CenterforHighEnergyPhysicsKyungpookNationalDaegu702701KoreaSeoulNationalSeoul151742KoreaSungkyunkwanSuwon440746KoreaKoreaScienceandTechnologyInformationDaejeon305806KoreaChonnamNationalGwangju500757Korea]{J.S.~Suh}
\author[FloridaGainesvilleFlorida32611]{A.~Sukhanov}
\author[JointforNuclearResearchRU141980DubnaRussia]{I.~Suslov}
\author[TsukubaTsukubaIbaraki305Japan]{T.~Suzuki}
\author[IllinoisUrbanaIllinois61801]{A.~Taffard$^f$}
\author[OkayamaOkayama7008530Japan]{R.~Takashima}
\author[TsukubaTsukubaIbaraki305Japan]{Y.~Takeuchi}
\author[OkayamaOkayama7008530Japan]{R.~Tanaka}
\author[MichiganAnnArborMichigan48109]{M.~Tecchio}
\author[PhysicsAcademiaSinicaTaipeiTaiwan11529RepublicChina]{P.K.~Teng}
\author[TheRockefellerNewYorkNewYork10021]{K.~Terashi}
\author[FermiNationalAcceleratorLaboratoryBataviaIllinois60510]{J.~Thom$^h$}
\author[GlasgowGlasgowG128QQUnitedKingdom]{A.S.~Thompson}
\author[IllinoisUrbanaIllinois61801]{G.A.~Thompson}
\author[PennsylvaniaPhiladelphiaPennsylvania19104]{E.~Thomson}
\author[YaleNewHavenConnecticut06520]{P.~Tipton}
\author[CentrodeInvestigacionesEnergeticasMedioambientalesyTecnologicasE28040MadridSpain]{P.~Ttito-Guzm\'{a}n}
\author[FermiNationalAcceleratorLaboratoryBataviaIllinois60510]{S.~Tkaczyk}
\author[TexasAMCollegeStationTexas77843]{D.~Toback}
\author[Comenius84248BratislavaSlovakiaExperimentalPhysics04001KosiceSlovakia]{S.~Tokar}
\author[MichiganStateEastLansingMichigan48824]{K.~Tollefson}
\author[TsukubaTsukubaIbaraki305Japan]{T.~Tomura}
\author[FermiNationalAcceleratorLaboratoryBataviaIllinois60510]{D.~Tonelli}
\author[LaboratoriNazionalidiFrascatiIstitutoNazionalediFisicaNucleareI00044FrascatiItaly]{S.~Torre}
\author[FermiNationalAcceleratorLaboratoryBataviaIllinois60510]{D.~Torretta}
\author[IstitutoNazionalediFisicaNucleareTriesteUdineI34100TriesteeeTriesteUdineI33100UdineItaly]{P.~Totaro$^{ee}$}
\author[LPNHEUniversitePierreetMarieCurieIN2P3CNRSUMR7585ParisF75252France]{S.~Tourneur}
\author[IstitutoNazionalediFisicaNuclearePisaaaPisabbSienaandccScuolaNormaleSuperioreI56127PisaItaly]{M.~Trovato$^{cc}$}
\author[PhysicsAcademiaSinicaTaipeiTaiwan11529RepublicChina]{S.-Y.~Tsai}
\author[PennsylvaniaPhiladelphiaPennsylvania19104]{Y.~Tu}
\author[IstitutoNazionalediFisicaNuclearePisaaaPisabbSienaandccScuolaNormaleSuperioreI56127PisaItaly]{N.~Turini$^{bb}$}
\author[TsukubaTsukubaIbaraki305Japan]{F.~Ukegawa}
\author[GenevaCH1211Geneva4Switzerland]{S.~Vallecorsa}
\author[DivisionHighEnergyPhysicsDepartmentPhysicsHelsinkiandHelsinkiPhysicsFIN00014HelsinkiFinland]{N.~van~Remortel$^b$}
\author[MichiganAnnArborMichigan48109]{A.~Varganov}
\author[IstitutoNazionalediFisicaNuclearePisaaaPisabbSienaandccScuolaNormaleSuperioreI56127PisaItaly]{E.~Vataga$^{cc}$}
\author[FloridaGainesvilleFlorida32611]{F.~V\'{a}zquez$^n$}
\author[FermiNationalAcceleratorLaboratoryBataviaIllinois60510]{G.~Velev}
\author[Athens15771AthensGreece]{C.~Vellidis}
\author[CentrodeInvestigacionesEnergeticasMedioambientalesyTecnologicasE28040MadridSpain]{M.~Vidal}
\author[FermiNationalAcceleratorLaboratoryBataviaIllinois60510]{R.~Vidal}
\author[InstitutodeFisicadeCantabriaCSICCantabria39005SantanderSpain]{I.~Vila}
\author[InstitutodeFisicadeCantabriaCSICCantabria39005SantanderSpain]{R.~Vilar}
\author[CollegeLondonLondonWC1E6BTUnitedKingdom]{T.~Vine}
\author[NewMexicoAlbuquerqueNewMexico87131]{M.~Vogel}
\author[ErnestOrlandoLawrenceBerkeleyNationalLaboratoryBerkeleyCalifornia94720]{I.~Volobouev$^u$}
\author[IstitutoNazionalediFisicaNuclearePisaaaPisabbSienaandccScuolaNormaleSuperioreI56127PisaItaly]{G.~Volpi$^{aa}$}
\author[PennsylvaniaPhiladelphiaPennsylvania19104]{P.~Wagner}
\author[ArgonneNationalLaboratoryArgonneIllinois60439]{R.G.~Wagner}
\author[FermiNationalAcceleratorLaboratoryBataviaIllinois60510]{R.L.~Wagner}
\author[Institutf"urExperimentelleKernphysikUniversit"atKarlsruhe76128KarlsruheGermany]{W.~Wagner$^x$}
\author[Institutf"urExperimentelleKernphysikUniversit"atKarlsruhe76128KarlsruheGermany]{J.~Wagner-Kuhr}
\author[OsakaCityOsaka588Japan]{T.~Wakisaka}
\author[CaliforniaLosAngelesLosAngelesCalifornia90024]{R.~Wallny}
\author[PhysicsAcademiaSinicaTaipeiTaiwan11529RepublicChina]{S.M.~Wang}
\author[ParticlePhysicsMcGillMontrealQuebecCanadaH3A2T8SimonFraserBurnabyBritishColumbiaCanadaV5A1S6TorontoTorontoOntarioCanadaM5S1A7andTRIUMFVancouverBritishColumbiaCanadaV6T2A3]{A.~Warburton}
\author[CollegeLondonLondonWC1E6BTUnitedKingdom]{D.~Waters}
\author[TexasAMCollegeStationTexas77843]{M.~Weinberger}
\author[Institutf"urExperimentelleKernphysikUniversit"atKarlsruhe76128KarlsruheGermany]{J.~Weinelt}
\author[FermiNationalAcceleratorLaboratoryBataviaIllinois60510]{W.C.~Wester~III}
\author[TuftsMedfordMassachusetts02155]{B.~Whitehouse}
\author[PennsylvaniaPhiladelphiaPennsylvania19104]{D.~Whiteson$^f$}
\author[ArgonneNationalLaboratoryArgonneIllinois60439]{A.B.~Wicklund}
\author[FermiNationalAcceleratorLaboratoryBataviaIllinois60510]{E.~Wicklund}
\author[EnricoFermiChicagoChicagoIllinois60637]{S.~Wilbur}
\author[ParticlePhysicsMcGillMontrealQuebecCanadaH3A2T8SimonFraserBurnabyBritishColumbiaCanadaV5A1S6TorontoTorontoOntarioCanadaM5S1A7andTRIUMFVancouverBritishColumbiaCanadaV6T2A3]{G.~Williams}
\author[PennsylvaniaPhiladelphiaPennsylvania19104]{H.H.~Williams}
\author[FermiNationalAcceleratorLaboratoryBataviaIllinois60510]{P.~Wilson}
\author[TheOhioStateColumbusOhio43210]{B.L.~Winer}
\author[FermiNationalAcceleratorLaboratoryBataviaIllinois60510]{P.~Wittich$^h$}
\author[FermiNationalAcceleratorLaboratoryBataviaIllinois60510]{S.~Wolbers}
\author[EnricoFermiChicagoChicagoIllinois60637]{C.~Wolfe}
\author[MichiganAnnArborMichigan48109]{T.~Wright}
\author[GenevaCH1211Geneva4Switzerland]{X.~Wu}
\author[CaliforniaSanDiegoLaJollaCalifornia92093]{F.~W\"urthwein}
\author[MassachusettsTechnologyCambridgeMassachusetts02139]{S.~Xie}
\author[CaliforniaSanDiegoLaJollaCalifornia92093]{A.~Yagil}
\author[OsakaCityOsaka588Japan]{K.~Yamamoto}
\author[DukeDurhamNorthCarolina27708]{J.~Yamaoka}
\author[EnricoFermiChicagoChicagoIllinois60637]{U.K.~Yang$^q$}
\author[CenterforHighEnergyPhysicsKyungpookNationalDaegu702701KoreaSeoulNationalSeoul151742KoreaSungkyunkwanSuwon440746KoreaKoreaScienceandTechnologyInformationDaejeon305806KoreaChonnamNationalGwangju500757Korea]{Y.C.~Yang}
\author[ErnestOrlandoLawrenceBerkeleyNationalLaboratoryBerkeleyCalifornia94720]{W.M.~Yao}
\author[FermiNationalAcceleratorLaboratoryBataviaIllinois60510]{G.P.~Yeh}
\author[FermiNationalAcceleratorLaboratoryBataviaIllinois60510]{K.~Yi$^o$}
\author[FermiNationalAcceleratorLaboratoryBataviaIllinois60510]{J.~Yoh}
\author[WasedaTokyo169Japan]{K.~Yorita}
\author[OsakaCityOsaka588Japan]{T.~Yoshida$^m$}
\author[RochesterRochesterNewYork14627]{G.B.~Yu}
\author[CenterforHighEnergyPhysicsKyungpookNationalDaegu702701KoreaSeoulNationalSeoul151742KoreaSungkyunkwanSuwon440746KoreaKoreaScienceandTechnologyInformationDaejeon305806KoreaChonnamNationalGwangju500757Korea]{I.~Yu}
\author[FermiNationalAcceleratorLaboratoryBataviaIllinois60510]{S.S.~Yu}
\author[FermiNationalAcceleratorLaboratoryBataviaIllinois60510]{J.C.~Yun}
\author[IstitutoNazionalediFisicaNucleareSezionediRoma1ddSapienzaUniversitadiRomaI00185RomaItaly]{L.~Zanello$^{dd}$}
\author[IstitutoNazionalediFisicaNucleareTriesteUdineI34100TriesteeeTriesteUdineI33100UdineItaly]{A.~Zanetti}
\author[IllinoisUrbanaIllinois61801]{X.~Zhang}
\author[CaliforniaLosAngelesLosAngelesCalifornia90024]{Y.~Zheng$^d$}
\author[IstitutoNazionalediFisicaNucleareBolognayBolognaI40127BolognaItaly]{S.~Zucchelli$^y$,}

\address{{\rm(CDF Collaboration\footnote{With visitors from $^a$University of Massachusetts Amherst, Amherst, Massachusetts 01003,
$^b$Universiteit Antwerpen, B-2610 Antwerp, Belgium, 
$^c$University of Bristol, Bristol BS8 1TL, United Kingdom,
$^d$Chinese Academy of Sciences, Beijing 100864, China, 
$^e$Istituto Nazionale di Fisica Nucleare, Sezione di Cagliari, 09042 Monserrato (Cagliari), Italy,
$^f$University of California Irvine, Irvine, CA  92697, 
$^g$University of California Santa Cruz, Santa Cruz, CA  95064, 
$^h$Cornell University, Ithaca, NY  14853, 
$^i$University of Cyprus, Nicosia CY-1678, Cyprus, 
$^j$University College Dublin, Dublin 4, Ireland,
$^k$University of Edinburgh, Edinburgh EH9 3JZ, United Kingdom, 
$^l$University of Fukui, Fukui City, Fukui Prefecture, Japan 910-0017
$^m$Kinki University, Higashi-Osaka City, Japan 577-8502
$^n$Universidad Iberoamericana, Mexico D.F., Mexico,
$^o$University of Iowa, Iowa City, IA  52242,
$^p$Queen Mary, University of London, London, E1 4NS, England,
$^q$University of Manchester, Manchester M13 9PL, England, 
$^r$Nagasaki Institute of Applied Science, Nagasaki, Japan, 
$^s$University of Notre Dame, Notre Dame, IN 46556,
$^t$University de Oviedo, E-33007 Oviedo, Spain, 
$^u$Texas Tech University, Lubbock, TX  79609, 
$^v$IFIC(CSIC-Universitat de Valencia), 46071 Valencia, Spain,
$^w$University of Virginia, Charlottesville, VA  22904,
$^x$Bergische Universit\"at Wuppertal, 42097 Wuppertal, Germany,
$^{ff}$On leave from J.~Stefan Institute, Ljubljana, Slovenia, 
})}}

\address[PhysicsAcademiaSinicaTaipeiTaiwan11529RepublicChina]{Institute of Physics, Academia Sinica, Taipei, Taiwan 11529, Republic of China} 
\address[ArgonneNationalLaboratoryArgonneIllinois60439]{Argonne National Laboratory, Argonne, Illinois 60439} 
\address[Athens15771AthensGreece]{University of Athens, 157 71 Athens, Greece} 
\address[InstitutdeFisicadAltesEnergiesUniversitatAutonomadeBarcelonaE08193Bellaterra(Barcelona)Spain]{Institut de Fisica d'Altes Energies, Universitat Autonoma de Barcelona, E-08193, Bellaterra (Barcelona), Spain} 
\address[BaylorWacoTexas76798]{Baylor University, Waco, Texas  76798} 
\address[IstitutoNazionalediFisicaNucleareBolognayBolognaI40127BolognaItaly]{Istituto Nazionale di Fisica Nucleare Bologna, $^y$University of Bologna, I-40127 Bologna, Italy} 
\address[BrandeisWalthamMassachusetts02254]{Brandeis University, Waltham, Massachusetts 02254} 
\address[CaliforniaDavisDavisCalifornia95616]{University of California, Davis, Davis, California  95616} 
\address[CaliforniaLosAngelesLosAngelesCalifornia90024]{University of California, Los Angeles, Los Angeles, California  90024} 
\address[CaliforniaSanDiegoLaJollaCalifornia92093]{University of California, San Diego, La Jolla, California  92093} 
\address[CaliforniaSantaBarbaraSantaBarbaraCalifornia93106]{University of California, Santa Barbara, Santa Barbara, California 93106} 
\address[InstitutodeFisicadeCantabriaCSICCantabria39005SantanderSpain]{Instituto de Fisica de Cantabria, CSIC-University of Cantabria, 39005 Santander, Spain} 
\address[CarnegieMellonPittsburghPA15213]{Carnegie Mellon University, Pittsburgh, PA  15213} 
\address[EnricoFermiChicagoChicagoIllinois60637]{Enrico Fermi Institute, University of Chicago, Chicago, Illinois 60637}
\address[Comenius84248BratislavaSlovakiaExperimentalPhysics04001KosiceSlovakia]{Comenius University, 842 48 Bratislava, Slovakia; Institute of Experimental Physics, 040 01 Kosice, Slovakia} 
\address[JointforNuclearResearchRU141980DubnaRussia]{Joint Institute for Nuclear Research, RU-141980 Dubna, Russia} 
\address[DukeDurhamNorthCarolina27708]{Duke University, Durham, North Carolina  27708} 
\address[FermiNationalAcceleratorLaboratoryBataviaIllinois60510]{Fermi National Accelerator Laboratory, Batavia, Illinois 60510} 
\address[FloridaGainesvilleFlorida32611]{University of Florida, Gainesville, Florida  32611} 
\address[LaboratoriNazionalidiFrascatiIstitutoNazionalediFisicaNucleareI00044FrascatiItaly]{Laboratori Nazionali di Frascati, Istituto Nazionale di Fisica Nucleare, I-00044 Frascati, Italy} 
\address[GenevaCH1211Geneva4Switzerland]{University of Geneva, CH-1211 Geneva 4, Switzerland} 
\address[GlasgowGlasgowG128QQUnitedKingdom]{Glasgow University, Glasgow G12 8QQ, United Kingdom} 
\address[HarvardCambridgeMassachusetts02138]{Harvard University, Cambridge, Massachusetts 02138} 
\address[DivisionHighEnergyPhysicsDepartmentPhysicsHelsinkiandHelsinkiPhysicsFIN00014HelsinkiFinland]{Division of High Energy Physics, Department of Physics, University of Helsinki and Helsinki Institute of Physics, FIN-00014, Helsinki, Finland} 
\address[IllinoisUrbanaIllinois61801]{University of Illinois, Urbana, Illinois 61801} 
\address[TheJohnsHopkinsBaltimoreMaryland21218]{The Johns Hopkins University, Baltimore, Maryland 21218} 
\address[Institutf"urExperimentelleKernphysikUniversit"atKarlsruhe76128KarlsruheGermany]{Institut f\"{u}r Experimentelle Kernphysik, Universit\"{a}t Karlsruhe, 76128 Karlsruhe, Germany} 
\address[CenterforHighEnergyPhysicsKyungpookNationalDaegu702701KoreaSeoulNationalSeoul151742KoreaSungkyunkwanSuwon440746KoreaKoreaScienceandTechnologyInformationDaejeon305806KoreaChonnamNationalGwangju500757Korea]{Center for High Energy Physics: Kyungpook National University, Daegu 702-701, Korea; Seoul National University, Seoul 151-742, Korea; Sungkyunkwan University, Suwon 440-746, Korea; Korea Institute of Science and Technology Information, Daejeon, 305-806, Korea; Chonnam National University, Gwangju, 500-757, Korea} 
\address[ErnestOrlandoLawrenceBerkeleyNationalLaboratoryBerkeleyCalifornia94720]{Ernest Orlando Lawrence Berkeley National Laboratory, Berkeley, California 94720} 
\address[LiverpoolLiverpoolL697ZEUnitedKingdom]{University of Liverpool, Liverpool L69 7ZE, United Kingdom} 
\address[CollegeLondonLondonWC1E6BTUnitedKingdom]{University College London, London WC1E 6BT, United Kingdom} 
\address[CentrodeInvestigacionesEnergeticasMedioambientalesyTecnologicasE28040MadridSpain]{Centro de Investigaciones Energeticas Medioambientales y Tecnologicas, E-28040 Madrid, Spain} 
\address[MassachusettsTechnologyCambridgeMassachusetts02139]{Massachusetts Institute of Technology, Cambridge, Massachusetts  02139} 
\address[ParticlePhysicsMcGillMontrealQuebecCanadaH3A2T8SimonFraserBurnabyBritishColumbiaCanadaV5A1S6TorontoTorontoOntarioCanadaM5S1A7andTRIUMFVancouverBritishColumbiaCanadaV6T2A3]{Institute of Particle Physics: McGill University, Montr\'{e}al, Qu\'{e}bec, Canada H3A~2T8; Simon Fraser University, Burnaby, British Columbia, Canada V5A~1S6; University of Toronto, Toronto, Ontario, Canada M5S~1A7; and TRIUMF, Vancouver, British Columbia, Canada V6T~2A3} 
\address[MichiganAnnArborMichigan48109]{University of Michigan, Ann Arbor, Michigan 48109} 
\address[MichiganStateEastLansingMichigan48824]{Michigan State University, East Lansing, Michigan  48824}
\address[InstitutionforTheoreticalandExperimentalPhysicsITEPMoscow117259Russia]{Institution for Theoretical and Experimental Physics, ITEP, Moscow 117259, Russia} 
\address[NewMexicoAlbuquerqueNewMexico87131]{University of New Mexico, Albuquerque, New Mexico 87131} 
\address[NorthwesternEvanstonIllinois60208]{Northwestern University, Evanston, Illinois  60208} 
\address[TheOhioStateColumbusOhio43210]{The Ohio State University, Columbus, Ohio  43210} 
\address[OkayamaOkayama7008530Japan]{Okayama University, Okayama 700-8530, Japan} 
\address[OsakaCityOsaka588Japan]{Osaka City University, Osaka 588, Japan} 
\address[OxfordOxfordOX13RHUnitedKingdom]{University of Oxford, Oxford OX1 3RH, United Kingdom} 
\address[IstitutoNazionalediFisicaNucleareSezionediPadovaTrentozPadovaI35131PadovaItaly]{Istituto Nazionale di Fisica Nucleare, Sezione di Padova-Trento, $^z$University of Padova, I-35131 Padova, Italy} 
\address[LPNHEUniversitePierreetMarieCurieIN2P3CNRSUMR7585ParisF75252France]{LPNHE, Universite Pierre et Marie Curie/IN2P3-CNRS, UMR7585, Paris, F-75252 France} 
\address[PennsylvaniaPhiladelphiaPennsylvania19104]{University of Pennsylvania, Philadelphia, Pennsylvania 19104}
\address[IstitutoNazionalediFisicaNuclearePisaaaPisabbSienaandccScuolaNormaleSuperioreI56127PisaItaly]{Istituto Nazionale di Fisica Nucleare Pisa, $^{aa}$University of Pisa, $^{bb}$University of Siena and $^{cc}$Scuola Normale Superiore, I-56127 Pisa, Italy} 
\address[PittsburghPittsburghPennsylvania15260]{University of Pittsburgh, Pittsburgh, Pennsylvania 15260} 
\address[PurdueWestLafayetteIndiana47907]{Purdue University, West Lafayette, Indiana 47907} 
\address[RochesterRochesterNewYork14627]{University of Rochester, Rochester, New York 14627} 
\address[TheRockefellerNewYorkNewYork10021]{The Rockefeller University, New York, New York 10021} 
\address[IstitutoNazionalediFisicaNucleareSezionediRoma1ddSapienzaUniversitadiRomaI00185RomaItaly]{Istituto Nazionale di Fisica Nucleare, Sezione di Roma 1, $^{dd}$Sapienza Universit\`{a} di Roma, I-00185 Roma, Italy} 
\address[RutgersPiscatawayNewJersey08855]{Rutgers University, Piscataway, New Jersey 08855} 
\address[TexasAMCollegeStationTexas77843]{Texas A\&M University, College Station, Texas 77843} 
\address[IstitutoNazionalediFisicaNucleareTriesteUdineI34100TriesteeeTriesteUdineI33100UdineItaly]{Istituto Nazionale di Fisica Nucleare Trieste/Udine, I-34100 Trieste, $^{ee}$University of Trieste/Udine, I-33100 Udine, Italy} 
\address[TsukubaTsukubaIbaraki305Japan]{University of Tsukuba, Tsukuba, Ibaraki 305, Japan} 
\address[TuftsMedfordMassachusetts02155]{Tufts University, Medford, Massachusetts 02155} 
\address[WasedaTokyo169Japan]{Waseda University, Tokyo 169, Japan} 
\address[WayneStateDetroitMichigan48201]{Wayne State University, Detroit, Michigan  48201} 
\address[WisconsinMadisonWisconsin53706]{University of Wisconsin, Madison, Wisconsin 53706} 
\address[YaleNewHavenConnecticut06520]{Yale University, New Haven, Connecticut 06520}

\begin{abstract}

Fully reconstructed $t\bar{t}\rightarrow W^{+}bW^{-}\bar{b} \rightarrow \ell\nu q \bar{q}^{\prime} b \bar{b}$ 
events are used to determine the fractions of right-handed ($f_{+}$) and longitudinally
polarized ($f_{0}$) $W$ bosons produced in
top-quark decays. The helicity fractions are sensitive to the
couplings and the Dirac structure of the $Wtb$ vertex.
This paper reports measurements of the $W$-boson helicity fractions from
two different methods
using data corresponding to an integrated luminosity of $1.9\,\mathrm{fb}^{-1}$ of $p\bar{p}$ collisions at a center-of-mass energy of 1.96~TeV collected by the CDF~II 
detector operating at the Fermilab Tevatron.  
Combining the results from the two methods, we find $f_{0} = 0.62\pm0.10\:\rm{(stat)}\:\pm0.05\:\rm{(syst)}$ under the
assumption that $f_{+}=0$, and $f_{+} = -0.04\pm0.04\:\rm{(stat)}\:\pm0.03\:\rm{(syst)}$ with $f_{0}$ fixed
to the theoretically expected value of $0.70$.  Model-independent fits are
also performed and simultaneously 
determine $f_{0} = 0.66\pm0.16\:\rm{(stat)}\:\pm0.05\:\rm{(syst)}$ and $f_{+} =-0.03\pm0.06\:\rm{(stat)}\:\pm0.03\:\rm{(syst)}$.
All these results are consistent with standard model expectations.
\end{abstract}
\begin{keyword}
   helicity \sep W boson \sep top quark

   \PACS 12.15.-y \sep 13.38.Be \sep 13.88.+e \sep 14.65.Ha \sep 14.70Fm
\end{keyword}
\end{frontmatter}







\section{Introduction}
\label{sec:introduction}
Charged current weak interactions proceed via the exchange of a $W^\pm$ boson
and are theoretically described by a vertex factor that has a pure vector minus axial-vector 
($V-A$) structure~\cite{v-a}. 
While weak interactions have been tested with high precision at low 
momentum transfers, e.g. in radioactive $\beta$-decay, the vertex structure may be
altered in interactions at high momentum transfers due to new physics
contributions.
Among the known fundamental particles, the top quark stands out as the
heaviest, with a mass of $m_{t} = 172.4\pm 1.2~\GeVc2$~\cite{c:mtop}, and thereby gives
access to high momentum scales. It has been suggested that
the top quark may have non-universal gauge couplings as a result of
dynamical breaking of the electroweak symmetry~\cite{peccei}.

Given our present knowledge of the Cabibbo-Kobayashi-Maskawa quark-mixing
matrix~\cite{c:PDG}, the top quark decays with a branching ratio close
to 100\% in the mode $t\rightarrow b W^+$.
The Dirac structure of the $Wtb$ vertex can be generalized by the
interaction Lagrangian

\begin{equation}
  \label{eq:Wtb}
  \mathcal{L} = \frac{g_w}{\sqrt{2}} \left[ W_\mu^- \bar{b}\gamma^\mu
  (f_1^L P_- + f_1^R P_+ ) t - \frac{1}{m_W} \partial_\nu W_\mu^- \bar{b}
  \sigma^{\mu\nu} (f_2^L P_- + f_2^R P_+) t \right] + h.c.\;,
\end{equation}
where $P_\pm = \frac{1}{2}(1\pm\gamma^5)$ and 
$i\sigma^{\mu\nu}=-\frac{1}{2}[\gamma^\mu, \gamma^\nu]$~\cite{c:kane}.
In general the interaction of fermions and gauge bosons can be expressed by six form factors.
Assuming the $W$ boson to be on-shell, the number of form factors is reduced to four.
These four form factors $f_{1,2}^{L,R}$ can assume complex values in general, 
but take values of $f_1^L=1$ and $f_1^R=f_2^L=f_2^R=0$ in standard electroweak 
theory, such that the production of right-handed $W$ bosons from top-quark
decay is suppressed. A general strategy to experimentally determine all four form factors in Eq.~\ref{eq:Wtb}
involves the measurement of the $W$-boson helicity fractions and the
measurement of the single top-quark production cross-section in the
$t$-channel and in the $s$-channel~\cite{chen}.

The production of longitudinally polarized $W$ bosons is
enhanced due to the large Yukawa coupling of the top quark to the
Higgs field responsible for electroweak symmetry breaking (EWSB).  
The fraction of right-handed $W$ bosons, $f_{+}$, is predicted to be very small
${\cal{O}}(10^{-4})$~\cite{c:fplus}, which is well below the sensitivity of the
measurements reported here. 
The partial decay widths into the different $W$-boson helicity states explicitly depend on the form factors.
Assuming the standard electroweak theory values for the form factors  
the fraction of longitudinally polarized $W$ bosons is  given by 
$f_{0} = \frac{\Gamma(W_{0})}{\Gamma(W_{0})+\Gamma(W_{-})+\Gamma(W_{+})}\approx {\frac{m^{2}_{t}}{2m^{2}_{W}+m^{2}_{t}}}$~\cite{c:kane} at leading order in perturbation theory, where $W_0$ and $W_\pm$ indicate longitudinally and
transversely polarized $W$ bosons, respectively. 
For $m_{t}$  as given above and a $W$-boson mass of 
$m_W = 80.403\pm 0.029~\GeVc2$~\cite{c:PDG} the theory predicts $f_{0} = 0.697\pm 0.002$.  
Next-to-leading-order corrections decrease the total decay width, 
as well as $\Gamma(W_{0})$, by about 10\%~\cite{kuehn},
while $f_{0}$ is only changed by about 1\%~\cite{koerner}.
A significant deviation of $f_{0}$ or $f_{+}$ from the predictions 
exceeding the 1\% level would be a clear indication of new 
physics.

This article reports the results of two analyses using the same dataset and their
combination. Both analyses use the observable $\cos\theta^{*}$, which is the cosine of the decay 
angle of the charged lepton in the $W$-boson decay frame measured with respect to 
the top-quark direction. This has the following distribution:
\begin{eqnarray}
\label{eq:omega}
\omega (\theta^{*}) = { f_{0} \cdot \omega_0 (\theta^{*}) +
                   { f_{+} \cdot \omega_+ (\theta^{*}) +
                   { (1-f_{0} -f_{+}) \cdot \omega_- (\theta^{*}) }}} 
  \ \ \mathrm{with} \\ 
\omega_0 (\theta^{*}) = \frac{3}{4}(1-\cos^{2}\theta^{*}), 
\omega_+ (\theta^{*}) = \frac{3}{8}(1+\cos\theta^{*})^{2},
\omega_- (\theta^{*}) = \frac{3}{8}(1-\cos\theta^{*})^{2}. 
\end{eqnarray}
\noindent
The parameters $f_{0}$ and $f_{+}$ are the $W$-boson helicity fractions to be determined.

The two analyses estimate $\cos\theta^{*}$ for each event by reconstructing the full 
$t\bar{t}$ kinematics. These methods of reconstructing the four-vectors of the
top-quark and antitop-quark as well as their decay 
products~\cite{c:CDFMtTMT,c:PHD_Dominic,c:Diploma_Thorsten} possess a broad 
applicability and
offer the possibility to measure a full set of top-quark properties, such as the 
top-quark mass and
the forward-backward charge asymmetry in $t\bar{t}$ production~\cite{c:AFB_CDF}.
Experimental acceptances and resolutions introduce
distortions of the $\cos\theta^{*}$ distribution which must also be taken into account.  The two analyses
employ alternative methods for reconstructing the $t\bar{t}$ kinematics, for
correcting the experimental effects, and for determining the polarization
fractions from the resulting $\cos\theta^{*}$ distributions in the observed events. They
have similar sensitivities and are combined, taking into account
correlations, to yield the most precise estimates of $f_{0}$ and $f_{+}$.
Both analyses subject the observed data to fits in three different scenarios:
\begin{enumerate}
\item Measure $f_0$ under the assumption that $f_+=0$. This corresponds to a
   model in which the form factors $f_1^R$ and $f_2^L$ are zero, meaning there are no right-handed bottom-quark
couplings present.
\item Measure $f_+$ under the assumption that $f_0=0.7$, which is sensitive to
   models with $f_2^L=f_2^R=0$, i.e. the presence of an additional $V+A$ current in top-quark decay,
but no additional magnetic couplings. Using the relation
   $f_+/f_- = (f_1^R/f_1^L)^2$ one can translate the measured helicity fractions
   into the ratio of form factors.
\item Measure $f_0$ and $f_+$ simultaneously in a two-parameter fit, which is
   model-independent.  
\end{enumerate}

Model-dependent measurements of $f_{0}$ and $f_{+}$ using smaller datasets
have been previously reported by the CDF~\cite{c:KarlPublication} and
\DZero~\cite{c:d0} collaborations.  Most recently the \DZero
collaboration has reported a model-independent result using
$1~\mathrm{fb}^{-1}$~\cite{c:d0} of Tevatron data.  The measurements reported
here use twice as much data and improved analysis techniques and yield
the most precise determinations of the $W$-boson helicity fractions in top-quark decays.


\section{Selection of \boldmath$t\bar{t}$ candidate events}
\label{sec:Eventselection}

The data used for the analyses reported here are collected by the CDF II
detector~\cite{c:cdfDet}.
We select events of the type $t\bar{t}\rightarrow W^{+}bW^{-}\bar{b} \rightarrow \ell\nu q \bar{q}^{\prime} b \bar{b}$,
which yield an experimental signature of one high energy charged
lepton, missing transverse energy due to the undetected neutrino, and
at least four jets, two of which are $b$-quark jets. 
Exactly one isolated electron candidate with transverse energy~\cite{c:geometry} $E_{\rm{T}}>20~\GeV$ and pseudorapidity~\cite{c:geometry} $|\eta|<1.1$ is
required, or exactly one isolated muon candidate with transverse momentum~\cite{c:geometry} $P_{\rm{T}}>20~\ensuremath{\mathrm{ Ge\kern -0.1em V }\kern -0.2em /c}$ and
$|\eta|<1.0$. An electron or muon candidate is considered isolated if
the $E_{\rm{T}}$ not assigned to the lepton in a cone of $R\equiv
\sqrt{(\Delta\eta)^{2}+(\Delta\phi)^{2}}=0.4$, centered around the
lepton, is less than 10\% of the lepton $E_{\rm{T}}$ or $P_{\rm{T}}$,
respectively. Jets are reconstructed by summing calorimeter energy in
a cone of radius $R=0.4$. The energy of the jets is corrected for differences as a function of
 $\eta$, time, and additional energy depositions due to multiple interactions occurring in the same event~\cite{c:jetEnergy}. An additional correction leads from calorimeter based jets to jets at the particle level.
Candidate jets must have corrected $E_{\rm{T}}>20~\GeV$ and detector $|\eta| <
2$. Events are
required to have at least four jets.  The corrected missing transverse energy~\cite{c:MET} \MET\ accounts
for the energy corrections made for all jets with corrected
$E_{\rm{T}}>12~\GeV$ and $|\eta|<2.4$ and for muons and is required to be greater than $20~\GeV$.
 At least one jet in the event has to contain a secondary vertex identified using
the algorithm described in~\cite{c:xsec} and consistent with
having originated from a $b$-hadron decay.  Additional requirements 
further reduce
the background contribution as follows. Electron events are rejected if the
electrons originate from a conversion of a photon. Cosmic ray muon
events are rejected as well. To remove $Z$ bosons, events in
which the charged lepton can be paired with any more loosely defined
jet or lepton to form an invariant mass consistent with the $Z$ peak,
$76-106~\GeVc2$, are excluded. With these selection criteria, we
select 484 $t\bar{t}$ candidates in a sample corresponding to a total
integrated luminosity of $1.9~\mathrm{fb}^{-1}$.

Kinematic resolutions and selection and reconstruction efficiencies
for  $t\bar{t}$ events are determined utilizing \textsc{pythia}~\cite{c:pythia} and 
\textsc{herwig}~\cite{c:herwig}~event generators where the top-quark mass is set to $175~\GeVc2$. Samples of events generated with  \textsc{pythia},
 \textsc{alpgen}~\cite{c:alpgen}, and  \textsc{madevent}~\cite{c:madevent}, interfaced to \textsc{pythia} parton showering are used to determine certain background rates
and to estimate the  $\cos\theta^{*}$ distribution for background events.
In order to develop and validate the methods presented, \textsc{madevent} and a custom version of \textsc{herwig} are used to generate samples of simulated events with controllable $W$-boson helicity fractions.
  All generated events are passed through
the CDF detector simulation~\cite{c:cdfSim} and then reconstructed in
the same way as the observed events.


\section{Background Estimation}

The selected $t\bar{t}$\ sample is estimated to be contaminated with about 87 events
coming from background processes. These non-$t\bar{t}$ processes
originate mainly from $W$+jets events with a falsely reconstructed
secondary vertex (Mistags), from $W$+jets events in which the jets are
real $b$- and $c$-quark jets ($W$+heavy flavor), and multi-jet processes that contain no real $W$~boson (non-$W$). These backgrounds are
estimated using a combination of data and Monte Carlo methods as
described in detail in~\cite{c:xsec}.  Additional sources of
background arise from electroweak processes like diboson production ($WW$, $WZ$, $ZZ$), the production of 
single top-quarks, and $Z$ bosons. These backgrounds are predicted based on their theoretical cross sections and acceptances and efficiencies, which are derived from simulated events.
Table~\ref{t:bgd} shows the background estimation and the
observed number of events after all selection criteria.

\begin{table}[tbh]
\begin{center}
\begin{tabular}{lr}\hline\hline
   Background source    & N($\geq4$ jet) \\ \hline
   $W$+heavy flavor       & $37 \pm 10$\\
   Mistags              & $20 \pm 5$\\
   non-$W$              & $18 \pm 16$\\ 
   Electroweak           & $12 \pm 1$\\\hline
   Total background     & $87 \pm 23$\\ 
   Observed events      & $484$ \\ \hline\hline
\end{tabular}
\caption{\label{t:bgd} \small{Expected number of background events and
   the number of observed events in a $1.9~\mathrm{fb}^{-1}$ data sample using the selection criteria described
   in the text.}}
\end{center}
\end{table}


\section{Extraction of the \boldmath$W$-boson helicity fractions}

In order to measure the $W$-boson helicity fractions we follow two approaches. Both analyses use $\cos\theta^{*}$ as the sensitive observable, estimated on 
an event-by-event basis by fully reconstructing the $t\bar{t}$ kinematics.
The $\cos\theta^{*}$ distribution can be decomposed into three separate components according to the three different $W$-boson helicity states.
The first analysis is based on the methods developed to precisely measure the top-quark mass~\cite{c:CDFMtTMT} and uses the fact that the three helicity components
have distinguishable shapes. In this technique we find the expected distributions (``templates'') of the helicity components, containing resolution effects, and superpose those. The helicity fractions are 
then given by normalizations from an unbinned likelihood fit and the results are corrected for acceptance effects afterwards~\cite{c:Shulamit}.
We refer to this analysis as the ``template analysis'' in the following.
The second analysis, called the ``convolution analysis'', is based on the method described in~\cite{ c:PHD_Dominic, c:Diploma_Thorsten, c:KarlPublication}.
Starting from the theoretically predicted number of events in each bin of the particle level $\cos\theta^{*}$ distribution we convolute acceptance and resolution effects with these predictions
to derive the expected number of events in each bin of the reconstructed $\cos\theta^{*}$ distribution.
In this method, $f_{0}$ and $f_{+}$ are then determined from a binned likelihood fit.

The event selection and reconstruction of the two techniques employ different choices in the design of background suppression, jet flavor identification, and parton assignment.
The agreement between the two methods shows that these design choices do not bias the final result.
While the convolution analysis uses the standard event selection described in sec.~\ref{sec:Eventselection}, the template analysis chooses to place an additional cut on the scalar sum of all transverse energies 
of the event, $H_{\rm{T}}$, and requires $H_{\rm{T}}~>~250~\GeV$ to further suppress multi-jet non-$W$ background. This results in $53~\pm~20$ events estimated as background, and reduces 
the total number of selected events to 430.
A combinatoric ambiguity arises in the reconstruction of the $t\bar{t}$ kinematics when choosing which of the reconstructed jets corresponds to
which of the final state quarks in the $t\bar{t}\rightarrow\ell\nu q \bar{q}^{\prime} b \bar{b}$ decay.  The analyses
each test all possible jet-quark assignments and then use alternative criteria to choose the ``best'' one for each event.
The template analysis uses the technique described in~\cite{c:CDFMtTMT}: jet energies float within expected resolutions, $b$-tagged jets are assigned to $b$ quarks,
and the top-quark mass is left floating in the fit while the $W$-boson masses are constrained to $80.4~\GeVc2$.
The algorithm described in~\cite{c:PHD_Dominic, c:Diploma_Thorsten, c:KarlPublication} is used in the convolution analysis. The jet-quark assignment is selected using constraints on the $W$-boson mass, the 
$t\bar{t}$ mass difference, the transverse energy in the reconstructed $t\bar{t}$ pair with respect to the total transverse energy in the event, and the
$b$-jet probability of the jets.
Neither analysis assumes a particular value for $m_{t}$ in the reconstruction; since $f_{0}$ has
an explicit $m_{t}$-dependence, doing so would introduce a bias in the measurement.
Although the algorithms to reconstruct the kinematics of the $t\bar{t}$ pairs are different, the $\cos\theta^{*}$ resolution for each analysis
is estimated to be the same ($\approx$ 0.35) from studies using generated $t\bar{t}$ events.

In both analyses the $W$-boson helicity fractions are determined from maximum likelihood fits
to the resulting $\cos\theta^{*}$ distributions. 
The two analyses employ alternative methods to derive the fit inputs which will be discussed in more detail in the next paragraphs.
In the fits, the helicity fractions $f_{0}$ and $f_{+}$ are free parameters, the constraint $f_{-} =
(1-f_{0}-f_{+})$ is applied, and the background contribution is allowed to
float but is Gaussian constrained using an RMS corresponding to the
uncertainty on the estimate of the total number of background events.
As already discussed in sec.~\ref{sec:introduction}, each analysis performs three different measurements. In two measurements we determine $f_{0}$ or $f_{+}$ and fix the other
parameter to the value expected in case of a pure $V-A$ structure of the $Wtb$ vertex ($f_{0}$ = 0.7, $f_{+}$ = 0.0).
In the third measurement, $f_{0}$ and $f_{+}$ are both treated as free parameters and are measured simultaneously.

\begin{figure}[t]
\begin{center}
\includegraphics[width=0.5\columnwidth]{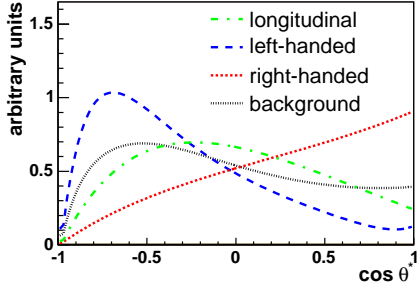}
\caption{ \small {The $\mathcal{P}(\cos{\theta^{*}})$ used
in the template analysis, which are the reconstructed $\omega
(\theta^{*})$ distributions for longitudinal, right- and left-handed
$W$-boson helicities, as well as the $\omega (\theta^{*})$ in the
background model. The curves are normalized to the same area.}
\label{fig:pdfs}}
\end{center}
\end{figure}

The template method utilizes samples of generated $t\bar{t}$ events in which the
leptonically decaying $W$ boson is forced to a specific polarization to get
the normalized $\cos\theta^{*}$ probability distribution function $\mathcal{P}(\cos{\theta^{*}})$ for each $W$-boson polarization.
 These generated events satisfy all the selection criteria and are reconstructed in the same manner as the observed events.
The $\mathcal{P}(\cos{\theta^{*}})$ for a certain helicity mode is obtained by fitting the reconstructed $\cos\theta^{*}$ distribution obtained
from the corresponding generated  $t\bar{t}$ events and does not depend on the
helicity fractions assumed for the hadronically decaying $W$ boson. 
The background modeling is verified by comparing
the distribution obtained from generated events to the distribution of observed events in which there is no secondary vertex
tag and in those for which the decay length of the secondary vertex
tag is negative, meaning that the reconstructed secondary vertex and the reconstructed jet itself are located in opposite hemispheres
 with respect to the primary vertex.  These are background dominated samples.  The $\mathcal{P}(\cos{\theta^{*}})$ parameterizations are 
empirically chosen to provide a good description
of the $\cos\theta^{*}$ distributions and use a third degree polynomial times two
exponential functions.  The resulting $\mathcal{P}(\cos{\theta^{*}})$ are compared in
Fig.~\ref{fig:pdfs}. Using alternative fit functions, negligibly affects the results.

Since the kinematics of the $W$-boson decay depend on its polarization,
the kinematic cuts applied have different acceptances for the different
polarizations and alter the observed composition of polarization states.
The largest impact is due to the isolation requirement and the cut on the $p_{T}$ of the charged lepton.
Therefore a correction is applied to the obtained helicity fractions to
account for these acceptance effects before presenting the results.

In the convolution analysis the $\cos\theta^{*}$ distribution is reconstructed in six bins, corresponding to the resolution of the reconstruction
of the $t\bar{t}$ kinematics. The starting point for the extraction of the $W$-boson helicity fractions in this method is the theoretically predicted
number of signal events in each bin of the $\cos\theta^{*}$ distribution, $\mu^{\rm{sig}}(f_{0},f_{+})$, depending on $f_{0}$ and $f_{+}$, which can be calculated using Eq.~\ref{eq:omega}.
 Acceptance and resolution effects are then taken into account~\cite{c:KarlPublication} by convoluting 
both effects with the theory prediction. This leads to the number of signal events expected to be observed in a certain bin accounting for all distorting effects:
\begin{eqnarray}
\label{eq:mu}
\mu_{k}^{\rm{sig,obs}}(f_{0},f_{+}) \propto \sum_{i}\mu_{i}^{\rm{sig}}(f_{0},f_{+})\cdot\epsilon_{i}\cdot S(i,k) .
\end{eqnarray}
\noindent

The migration matrix element $S(i,k)$ gives the probability for an event which was generated in bin $i$ to occur in bin $k$ of the
reconstructed $\cos{\theta^{*}}$ distribution. Since the acceptance depends on $\cos{\theta^{*}}$, we weight the contribution of each bin with its
event selection efficiency $\epsilon_{i}$. 
 The effects considered are independent of the $W$-boson helicity fractions and this is validated using several samples of generated events with
different $W$-boson polarizations. Thus, $\epsilon_{i}$ and $S(i,k)$ can be estimated from a sample of events generated with the \textsc{pythia} event generator using the standard settings.
The total number of events expected to be observed in a certain bin is then given by the sum of $\mu_{k}^{\rm{sig,obs}}(f_{0},f_{+})$ and the expected
number of background events, which is independent of the $W$-boson polarization and is derived from the background composition shown in 
Table~\ref{t:bgd}.
In a maximum-likelihood fit the expected number of events is compared bin by bin to the number of observed events to determine $f_{0}$ and $f_{+}$.

In order to compare our observations with theory, we subtract the background estimate from the reconstructed $\cos{\theta^{*}}$ distribution,
correct for acceptance and resolution, and normalize the distribution to the $t\bar{t}$ cross section of $\sigma_{t\bar{t}}=6.7\pm 0.9$~pb~\cite{c:Crosssection1,c:Crosssection2}.
The correction is made by applying a bin-by-bin correction factor to the $\cos\theta^{*}$ distribution. The correction factor is given by $\mu_{i}^{\rm{sig}}(f_{0}^{\rm{fit}},f_{+}^{\rm{fit}})$
 divided by $\mu_{i}^{\rm{sig,obs}}(f_{0}^{\rm{fit}},f_{+}^{\rm{fit}})$, where $f_{0}^{\rm{fit}}$ and $f_{+}^{\rm{fit}}$ are the obtained results.

\begin{table}[t]
  \begin{center}
    \begin{tabular}{lcccccccc}\hline \hline
    source &  \multicolumn{2}{c}{$\delta f_{0}$}& \multicolumn{2}{c}{$\delta f_{+}$} & \multicolumn{2}{c}{$\delta f_{0}$} & \multicolumn{2}{c}{$\delta f_{+}$}\\
        & \multicolumn{2}{c}{$f_{+}$ fixed} & \multicolumn{2}{c}{$f_{0}$ fixed} &  \multicolumn{2}{c}{combined fit} &  \multicolumn{2}{c}{combined fit}\\ 
        & templ. & conv. & templ. & conv. & templ. & conv. & templ. & conv. \\\hline
       JES & 0.024 & 0.045 & 0.017 & 0.025 & 0.021 & 0.016 & 0.027 & 0.032\\
       ISR & 0.002 & 0.010 & 0.003 & 0.003 & 0.010 & 0.036 & 0.007 & 0.014 \\
       FSR & 0.021 & 0.025 & 0.009 & 0.011 & 0.025 & 0.045 & 0.002 & 0.016 \\
       Bkg  & 0.023 & 0.032 & 0.016 & 0.019 & 0.018 & 0.028 & 0.017 & 0.032 \\	
       MC & 0.019 & 0.012 & 0.009 & 0.005 & 0.019 & 0.015 & 0.010 & 0.002 \\
       PDF & 0.005 & 0.005 & 0.005 & 0.002 & 0.005 & 0.014 & 0.002 & 0.006\\\hline
       Total  & 0.044 & 0.062 & 0.027 & 0.034  & 0.043 & 0.072 & 0.034 & 0.050 \\\hline \hline
          \end{tabular}
\caption{\label{t:sysUnfold} \small {The sources of systematic uncertainties
  and their related estimates for the template analysis (templ.) and the convolution analysis (conv.).  The total
  systematic uncertainty is taken as the quadrature sum of the
  individual sources.}}
    \label{tab:systematics}
  \end{center}
\end{table}

The systematic uncertainties associated with the measurement of
$f_{0}$ and $f_{+}$ are summarized in Table~\ref{tab:systematics}. The systematic
uncertainties were determined by constructing ensemble tests with
signal and/or background templates, affected by the systematic under
study, but fit using the default parameterizations and normalizations
described above. We studied the influence of variations in the jet energy scale (JES) and of variations in initial and final state radiation (ISR, FSR).
The latter was estimated by producing samples of simulated events for which the simulation was altered to produce either less or more gluon radiation compared to the standard setting~\cite{c:CDFMtTMT}. Specifically, two parameters controlling the parton shower in the
\textsc{pythia} program are varied: $\Lambda_\mathrm{QCD}$ and the
scale factor $K$ to the transverse momentum scale of the showering. 
The different settings are derived from studies of ISR in Drell-Yan events.
We also studied the influence of the background modeling (Bkg), of different Monte Carlo event generators (MC), and of the parton distribution function (PDF). The resulting shifts in the mean fitted longitudinal
and right-handed fraction are used to quantify the systematic
uncertainties. The positive and negative variations obtained are symmetrized by choosing the maximum deviation. The ensemble tests were all performed using \textsc{pythia}
generated events with $m_t = 175\:\GeVc2$ as signal with the
 $W$-boson helicity fractions $f_{0} = 0.70$ and $f_{+} = 0.0$, and
the background model as described above. We have verified that these
uncertainties do not depend on the actual value of $f_{0}$ and $f_{+}$ by
fitting samples of generated events with different $W$-boson polarizations.

The analyses presented in this paper use a top-quark
mass of 175~\GeVc2. Since $f_{0}$ explicitly depends on the top-quark mass, the dependency of the measured value of $f_{0}$ on the top quark mass is not treated as a systematic uncertainty. The measured value of $f_{+}$ is only negligibly affected by variations in the assumed top-quark mass.

\section{Results and Combination of the Results}

\begin{figure}[t]
\begin{center}
\includegraphics[width=0.5\columnwidth]{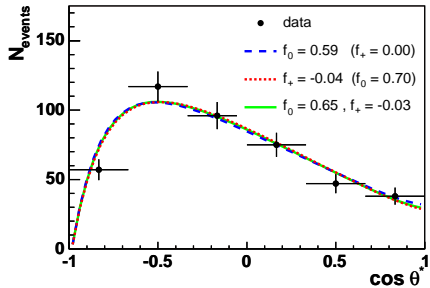}
\caption{ \small {The observed $\cos\theta^{*}$ distribution (points)
overlaid with the fit-curves for the three different fit-scenarios 
 (as explained in sec.~\ref{sec:introduction}) for the template analysis.}
\label{fig:resFNAL}}
\end{center}
\end{figure}

\begin{figure}[t]
\begin{center}
\includegraphics[width=0.5\columnwidth]{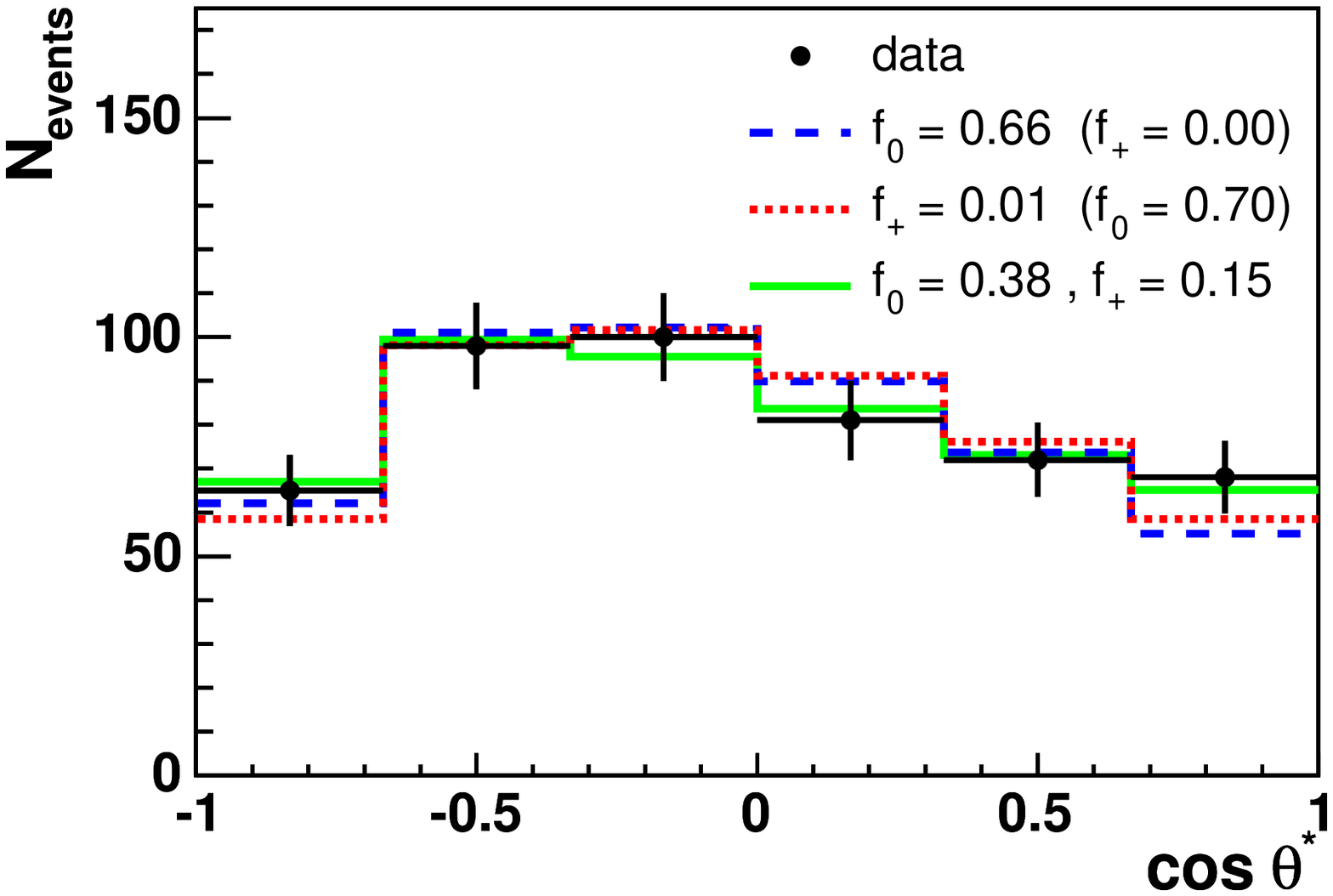}\includegraphics[width=0.5\columnwidth]{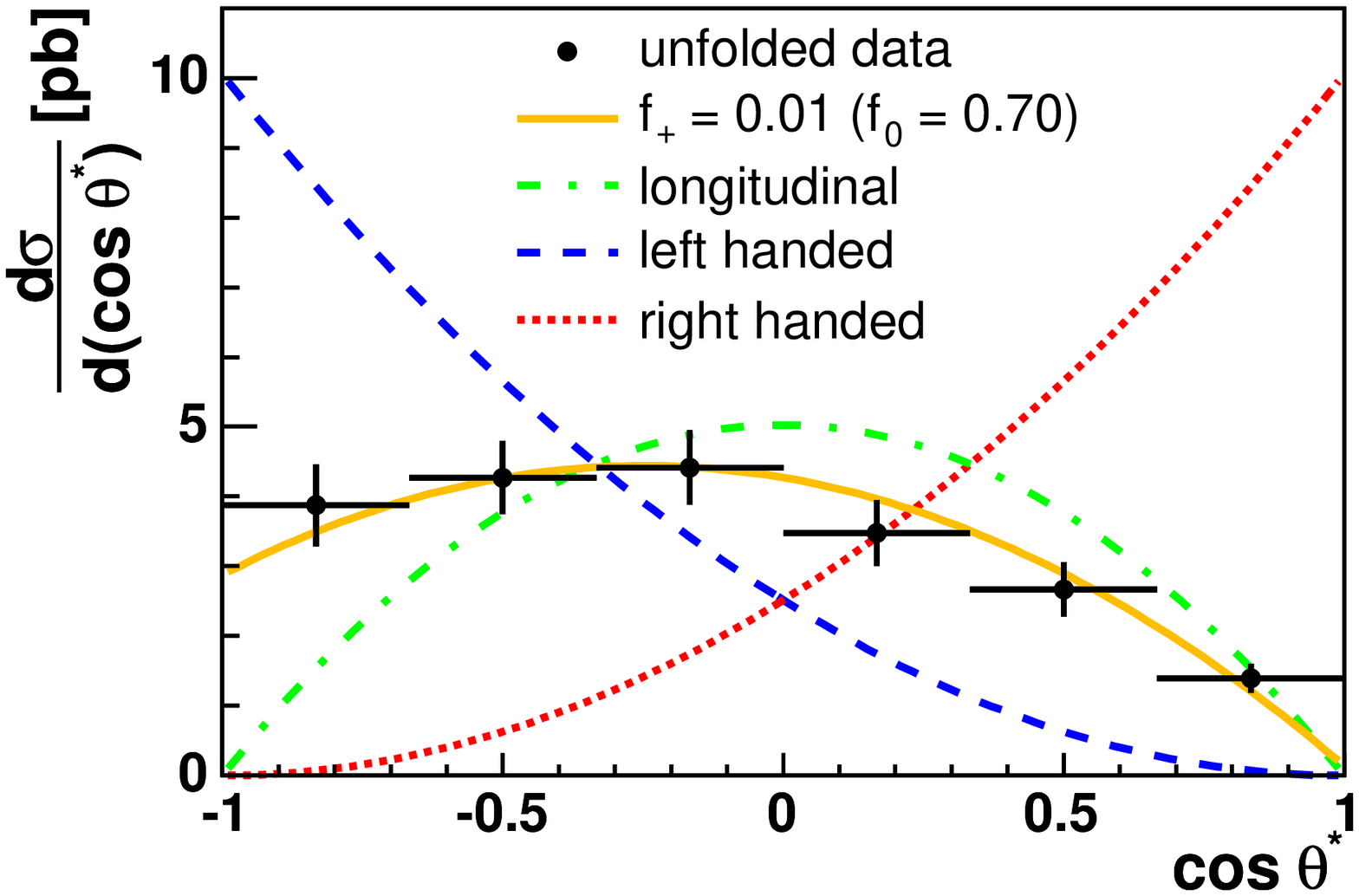}
\caption{ \small {On the left-hand side the observed $\cos\theta^{*}$ distribution (points)
is presented overlaid with the fits for $f_{0}$ and $f_{+}$ for the convolution analysis. On the right-hand side the deconvoluted (using the fit result of the $f_{+}$ measurement) distribution normalized to the $t\bar{t}$ cross-section is shown together with the theoretically predicted curves for purely left-handed, right-handed, and longitudinally polarized $W$ bosons.}
\label{fig:resKA}}
\end{center}
\end{figure}

The $\cos\theta^{*}$ distribution from the observed events is shown in
Fig.~\ref{fig:resFNAL} and Fig.~\ref{fig:resKA} for both analyses together with the fits for $f_{0}$ and $f_{+}$ and the model independent measurement. 
The results for the three different measurements together with the statistical and systematical uncertainties in both analyses are summarized in Table~\ref{t:results}.
In the template analysis the correlation between $f_{0}$ and $f_{+}$ is determined to be -0.87 in the simultaneous fit, while for the convolution analysis the correlation is -0.89.

\begin{table}[b]
\begin{center}
\begin{tabular}{l@{\hspace{0.3cm}} c@{\hspace{0.3cm}} c@{\hspace{0.3cm}} c@{\hspace{0.3cm}} c}\hline\hline
                        & template  & convolution & combination & $\chi^{2}/dof$  \\\hline
   $f_{0}(f_{+} = 0.0)$ & $0.59\pm0.11\pm0.04$ & $0.66\pm0.10\pm0.06$ & $0.62\pm0.10\pm0.05$ & 0.7/1 \\
   $f_{+}(f_{0} = 0.7)$ & $-0.04\pm0.04\pm0.03$ & $0.01\pm0.05\pm0.03$ &$-0.04\pm0.04\pm0.03$ & 1.8/1\\
   $f_{0}$              & $0.65\pm0.19\pm0.04$ & $0.38\pm0.21\pm0.07$ &$0.66\pm0.16\pm0.05$ & 4.3/2\\
   $f_{+}$              & $-0.03\pm0.07\pm0.03$ & $0.15\pm0.10\pm0.05$ &$-0.03\pm0.06\pm0.03$ & 4.3/2 \\\hline\hline

\end{tabular}
\caption{\label{t:results} \small{Results of the template analysis, the convolution analysis, and the combined values. The results are given together with their statistical and systematical uncertainties. In addition the $\chi^{2}/dof$ of the combination is given.}}
\end{center}
\end{table}

We combine the single results accounting for correlations using the BLUE method~\cite{c:blue}.
The combined results can be found in Table~\ref{t:results}.
The statistical correlation between both analyses is estimated from ensemble 
tests using samples of generated events which account for the event overlap in 
the signal contribution. 
For the two model-dependent scenarios the correlation coefficients are found to be 0.66
and 0.65 when fitting for $f_{0}$ or $f_{+}$, respectively.
The correlation matrix for the model-independent scenario is given in Table~\ref{t:corr}.
The resulting combination is weighted towards the template determination of $f_{+}$ since its total uncertainty is significantly smaller than 
the total uncertainty from the convolution method. Due to the strong anti-correlation between $f_{0}$ and $f_{+}$ (see Table~\ref{t:corr}) the $f_{0}$
determination is affected correspondingly.
The systematic uncertainties
are taken to be completely correlated between the two methods.
When combining the model-independent results the systematic uncertainties 
for $f_{0}$ and $f_{+}$ are taken to be 100\% anti-correlated. 
The combined values of $f_{0}$ and $f_{+}$ have a correlation of $-0.82$.
The combination improves the sensitivity by about $10\%$ relative to
the measurements of either method separately.

In conclusion, we present two different analyses and their combination
determining the $W$-boson helicity fractions in top-quark decays, giving the
world's most sensitive result for measuring these fractions so far.
In addition to measuring $f_{0}$ and $f_{+}$ separately, while fixing the other parameter to its expected value,
 we present a model-independent
simultaneous measurement of the two fractions. All of these results are consistent with the values predicted within the
electroweak theory of the $Wtb$ vertex.

\begin{table}[t!]
\begin{center}
\begin{tabular}{l@{\hspace{0.3cm}} c@{\hspace{0.3cm}} c@{\hspace{0.3cm}} c@{\hspace{0.3cm}} c}\hline\hline
                    & template $f_{0}$ & convolution $f_{0}$ & template $f_{+}$ & convolution $f_{+}$  \\\hline
 template $f_{0}$   & $1.00$  & $0.45$  & $-0.87$ & $-0.40$ \\
 convolution $f_{0}$& $0.45$  & $1.00$  & $-0.42$ & $-0.89$ \\
 template $f_{+}$   & $-0.87$ & $-0.42$ & $1.00$  & $0.48$  \\
 convolution $f_{+}$& $-0.40$ & $-0.89$ & $0.48$  & $1.00$  \\ \hline\hline
\end{tabular}
\caption{\label{t:corr} \small{Correlation matrix for combining the template and convolution 
  analyses in the model-independent scenario.}}
\end{center}
\end{table}

\section*{Acknowledgements}
We thank the Fermilab staff and the technical staffs of the participating institutions for their vital contributions. This work was supported by the U.S. Department of Energy and National Science Foundation; the Italian Istituto Nazionale di Fisica Nucleare; the Ministry of Education, Culture, Sports, Science and Technology of Japan; the Natural Sciences and Engineering Research Council of Canada; the National Science Council of the Republic of China; the Swiss National Science Foundation; the A.P. Sloan Foundation; the Bundesministerium f\"ur Bildung und Forschung, Germany; the Korean Science and Engineering Foundation and the Korean Research Foundation; the Science and Technology Facilities Council and the Royal Society, UK; the Institut National de Physique Nucleaire et Physique des Particules/CNRS; the Russian Foundation for Basic Research; the Ministerio de Ciencia e Innovaci\'{o}n, and Programa Consolider-Ingenio 2010, Spain; the Slovak R\&D Agency; and the Academy of Finland.


\begin{thebibliography}{99}
\bibitem{c:geometry}  In the CDF geometry,
$\theta$ is the polar angle with respect to the proton beam axis, and
$\phi$ is the azimuthal angle. The pseudorapidity is $\eta \equiv
-\ln(\tan(\theta/2))$. Detector $|\eta|$ is defined as the pseudorapidity of the jet
calculated with respect to the center of the detector. The transverse momentum $P_{\rm{T}}$ is the component of
the momentum projected onto the plane perpendicular to the beam
axis. The transverse energy $E_{\rm{T}}$ of a shower or calorimeter tower is
$E\sin\theta$, where $E$ is the energy deposited.
\bibitem{c:MET}  The missing $E_{\rm{T}}$ ($\vec{\MET}$) is defined by $\vec{\MET} =
-\Sigma_i {{E_{\rm{T}}^{i}}{\bf{n}}_{i}}$, where ${\bf{n}}_{i}$ is the unit vector in
the azimuthal plane that points from the beamline to the $i^{\rm{th}}$
calorimeter tower. We also define $ \MET = |\vec{\MET}|$.
\end{thebibliography}
\end{document}